\documentclass[aps,pra,twocolumn,superscriptaddress,showpacs]{revtex4}

\usepackage{graphicx}
\usepackage{tabularx}
\usepackage{amssymb}
\usepackage{amsmath}
\usepackage{bm}
\usepackage{subfigure}
\usepackage{graphpap}
\usepackage{multirow}
\usepackage{notes2bib}

\begin{document}
\title{Polymorphism in $\alpha$-sexithiophene crystals: Relative stability and transition path}
\author{Bernhard \surname{Klett}}
\affiliation{Institut f\"ur Physik, Humboldt-Universit\"at zu Berlin, Berlin, Germany}
\affiliation{IRIS Adlershof, Humboldt-Universit\"at zu Berlin, Berlin, Germany}
\author{Caterina \surname{Cocchi}}
\affiliation{Institut f\"ur Physik, Humboldt-Universit\"at zu Berlin, Berlin, Germany}
\affiliation{IRIS Adlershof, Humboldt-Universit\"at zu Berlin, Berlin, Germany}
\author{Linus \surname{Pithan}}
\affiliation{Institut f\"ur Physik, Humboldt-Universit\"at zu Berlin, Berlin, Germany}
\author{Stefan \surname{Kowarik}}
\affiliation{Institut f\"ur Physik, Humboldt-Universit\"at zu Berlin, Berlin, Germany}
\author{Claudia \surname{Draxl}}
\affiliation{Institut f\"ur Physik, Humboldt-Universit\"at zu Berlin, Berlin, Germany}
\affiliation{IRIS Adlershof, Humboldt-Universit\"at zu Berlin, Berlin, Germany}
\date{\today}

\begin{abstract}
We present a joint theoretical and experimental study to investigate polymorphism in $\alpha$-sexithiophene (6T) crystals.
By means of density-functional theory calculations, we clarify that the low-temperature phase is favorable over the high-temperature one, with higher relative stability by about 50 meV/molecule.
This result is in agreement with our thermal desorption measurements.
We also propose a transition path between the high- and low-temperature 6T polymorphs, estimating an upper bound for the energy barrier of about 1 eV/molecule. The analysis of the electronic properties of the investigated 6T crystal structures complements our study.
\end{abstract}

\maketitle
\section{Introduction}
\label{Sec: Introduction}
Organic crystalline semiconductors are promising materials for a variety of devices, ranging from light emitting diodes \cite{Forr2003OE, Forr2004NAT} to photovoltaics \cite{Peum-Yaki-Forr2003JAP}, and field-effect transistors \cite{Mucc2006NATM}. The possibility of synthesizing and processing these systems at low temperature and in solution is a particularly attractive feature \cite{Frax2006CUP}. For these reasons, organic crystals have attracted considerable interest in the last few decades. Depending on size and chemical composition of the molecular components as well as on the packing arrangement, it is possible to design systems with optimized properties.

Oligothiophenes represent a well-known family of organic crystals, which has been largely studied in view of opto-electronic applications \cite{dipp+93cpl,mark+95epl}. While the shortest thiophene oligomers are not suitable for device applications, due to their large band gap, most interest is devoted to longer chains such as $\alpha$-sexithiophene (6T). 
This material presents, in fact, a favorable combination of relatively high charge-carrier mobility \cite{horo+89ssc,akim+91apl,katz+95cm,Horo-Hajl-Kouk1998EPJ-AP} and visible light absorption \cite{oete+94jcp,oelk+96tsf,kane+96prb,fich00jmc,mark+98jpcb,Pith+2015CGD}, which makes it a very promising compound for organic electronics.

The weak interactions between 6T molecules enable the growth of two crystal structures, which are referred to as high- and low-temperature (HT and LT) phases \cite{Sieg+1995JMR, Horo+1995CM}. Both exhibit herringbone packing, with either two (HT) or four (LT) molecules per unit cell. Polymorphism, i.e., the presence of two or more possible arrangements of the molecules in the solid state \cite{desi08cgd}, is often observed in organic crystals. The coexistence of different morphologies of molecules in their crystal phases may strongly influence the properties of such materials \cite{bern93jpd,cair98tcc, lorch_growth_2015}. This has practical impact not only in condensed-matter physics and materials science \cite{thre95analyst,brag+98cr}, but also in biochemistry and pharmacology \cite{giro95thermca,rodr+04addr}, where polymorphism is known to crucially affect drug formulation and stability. Hence, a clear understanding of the fundamental mechanisms ruling this phenomenon is desired in view of tailoring molecular materials with customized features. 

In this paper, we address the question of polymorphism in 6T with a joint theoretical and experimental study. Specifically, we aim to determine which of the experimentally observed phases is the more stable one. To do so, we combine a first-principle approach, based on density-functional theory (DFT) and including van der Waals interactions, with thermal desorption measurements.
Moreover, we propose and analyze a possible transition path from one structure to the other, and estimate the size of the corresponding energy barrier. The information about the crystal structure is complemented by an analysis of the electronic properties of selected systems along the transition path.

The paper is organized as follows: In Sec. \ref{Sec: Systems and Methods}, we introduce the HT and LT polymorphs of 6T and the theoretical and experimental methods that we use. Sec. \ref{Sec: Relative Stability of HT and LT} adresses the relative stability of the two known polymorphs, both theoretically and experimentally. In Sec. \ref{Sec: Transition Path}, we discuss the transition path and the electronic properties corresponding to several structures along that path. 

\section{Systems and Methods}
\label{Sec: Systems and Methods}
\subsection{Sexithiophene crystals}
\label{Sec: Crystal Structure}

\begin{figure}
\centering
\includegraphics[width=.45\textwidth]{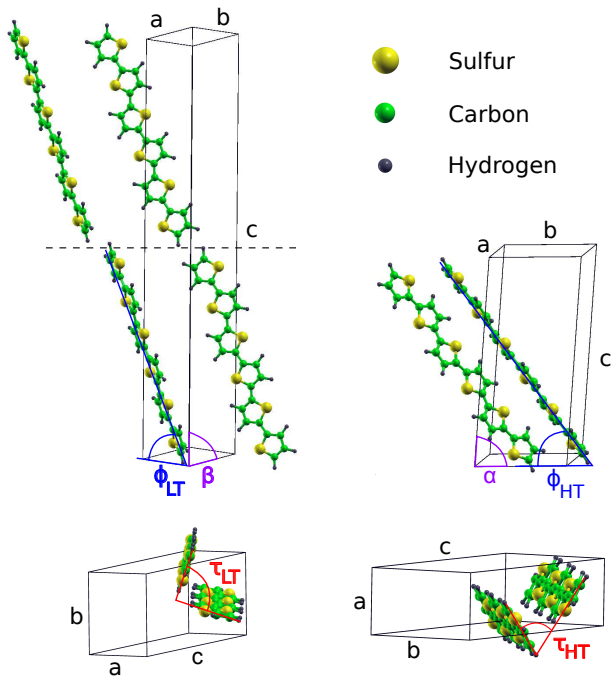}
\label{Structure-Notation} 
\caption{(color online) Unit cells of LT (left) and HT (right) polymorphs of 6T. Lattice paramters $a$, $b$, $c$, as well as tilt ($\phi$), herringbone ($\tau$), and monoclinic angles ($\alpha / \beta$) are indicated. The dashed line marks the plane, which divides the unit cell of the LT polymorph in half.\label{Fig: Crystal Structures}}
\end{figure}

\begin{table}
\setlength{\tabcolsep}{6pt}
\begin{tabular}{l r r r c c c}
\hline
& $a$ [\AA{}] & $b$ [\AA{}]& $c$ [\AA{}]& $\alpha$/$\beta$ [deg] & $\phi$ [deg] & $\tau$ [deg] \\
\hline \hline 
\textbf{HT} 	&\textbf{5.68}	&\textbf{9.14} &\textbf{20.67} & \textbf{97.8} & \textbf{48.5}	&\textbf{55.0} \\
\textrm{I}  &5.75	&8.88 &21.01 & 	&52.1	& 57.2 \\
\textrm{II}  &	5.79 &	8.75 &21.17 & &53.9 &	58.3\\
\textrm{III}  &5.82	&8.62 &	21.34	& &55.7 &	59.4 \\
\textrm{IV}  &	5.89 &	8.37 & 21.68 & &59.3	& 61.6 \\
\textrm{V}  &	5.93 &	8.24 &21.85 &	&61.1 &	62.7 \\
\textrm{VI} &	5.96 & 8.11 &22.01 & &62.9 &	63.8 \\
\textbf{LT} 	&\textbf{6.03}	&\textbf{7.85}	&\textbf{22.35}	& \textbf{90.8} & \textbf{66.5}	&\textbf{66.0} \\
\hline \hline
\end{tabular}
\caption{Structural parameters of the HT and LT polymorphs as well as of intermeditate structures (see Sec. \ref{Sec: Construction of the Path}). The lattice parameters of the LT polymorph are referred to the reduced unit cell (see Sec. \ref{Sec: Relative Stability of HT and LT}). Lattice constants $a$, $b$ and $c$, as well as the monoclinic angles $\alpha / \beta$, the tilt angle $\phi$, and the herringbone angle $\tau$ are displayed. \label{Tab:Structural Parameters}}
\end{table}

The building blocks of 6T crystals are planar molecules, consisting of carbon, hydrogen, and sulfur atoms, which are arranged in a chain of six rings. Sexithiophene exhibits a herringbone arrangement of the molecules in its crystal phases \citep{Sieg+1995JMR, Horo+1995CM}. The monoclinic unit cells of the LT and HT polymorphs are shown in Fig. \ref{Fig: Crystal Structures}. The corresponding structural parameters, based on x-ray diffraction experiments \citep{Sieg+1995JMR, Horo+1995CM}, are listed in Tab. \ref{Tab:Structural Parameters}. In the LT phase (left), the unit cell belongs to the space group $P2_1/n$, with a monoclinic angle $\beta_{LT}=90.8^\circ$ and contains four molecules, arranged in such a way, that the long molecular axes are almost parallel to each other. In this configuration, the tilt angle $\phi_{LT}=66.5^\circ$ is defined between the long molecular axis and the $ab$ plane, with lattice parameters $a=6.03\mbox{ \AA}$ and $b=7.85\mbox{ \AA}$. The herringbone angle $\tau_{LT}$ between the short molecular axes adopts a value of $66^\circ$ (bottom-left of Fig. \ref{Fig: Crystal Structures}). 

The HT structure is shown on the right-hand side of Fig. \ref{Fig: Crystal Structures}. Also in this case $a$ and $b$ have comparable values ($a=5.68$ \AA, $b=9.14$ \AA), while $c$ is much larger, being $20.67$ \AA. Note that we have permuted the crystal axes to highlight the similarity with the LT phase and to facilitate the construction of the transition path between them (see Sec. \ref{Sec: Transition Path}) \bibnote{Typically in crystallography the nomenclature for HT is such that the largest lattice parameter is labeled $a$ and the monoclinic angle is then $\beta$. Consequently, $b$ is the shortest lattice parameter. For the conventional nomenclature we refer to Ref. \cite{Sieg+1995JMR}.}. The space group for this structure is $P2_1/b$, with $\alpha_{HT}=97.8^\circ$. The volume of the HT unit cell is half as large as the LT one and accommodates two inequivalent molecules. Again, the long molecular axes are approximately parallel to each other. Both, the herringbone and tilt angle, that is defined with respect to the $ab$ plane, have smaller values than in the LT phase, being 55$^\circ$ and 48.5$^\circ$, respectively (see also Tab. \ref{Tab:Structural Parameters}).

\subsection{Computational details}
\label{Sec: Theory}
Total-energy and force calculations are performed within the framework of DFT.
All calculations are carried out with the full-potential all-electron code \texttt{exciting} \cite{Gula+2014JPCM},
implementing the (linearized) augmented planewave plus local orbitals method. 

A planewave cutoff $G_{max}\approx4.7 \mbox{ bohr}^{-1}$ is adopted for minimizing the atomic forces. This value corresponds to $R_{min}^{MT} G_{max}=3.5$, where $R_{min}^{MT}=0.75$ bohr is the smallest muffin-tin radius, corresponding to hydrogen. Radii of 1.15 bohr and 1.80 bohr are used for carbon and sulfur, respectively. To evaluate the small energy difference between the two polymorphs, we further increase the planewave cutoff to $G_{max}\approx6.7 \mbox{ bohr}^{-1}$, corresponding to $R_{min}^{MT} G_{max}=5.0$. We use $1 \times 5 \times 3$ and $4 \times 3 \times 1$ \textbf{k}-point grids for HT and LT, respectively. These parameters ensure the convergence of total energies within 0.4 meV/molecule.

For most calculations, we adopt the local density approximation (LDA), using the Perdew-Wang exchange-correlation functional \citep{LDA_PW}. The internal geometry is optimized by minimizing the atomic forces until they are smaller than 25 meV/\AA . The lattice parameters are thereby fixed at their experimental values (see Tab. \ref{Tab:Structural Parameters}). Doing so, the internal geometry depends largely on the electrostatic interactions. These are well described by LDA, as confirmed by the agreement of the interatomic distances with those reported in Ref. \cite{herm+05jpca}. Similar results have been found in an earlier study of anthracene \citep{Kerstin_Hummer}. 

To check the reliability of the LDA results and obtain more accurate energy differences between the two polymorphs, we also employ the DFT-D2 method \cite{Grim2006JCC} on top of the Perdew-Burke-Ernzerhof (PBE) exchange-correlation functional \citep{PBE} to calculate the total energies of the HT and LT polymorphs.

\subsection{Experimental methods}
\label{Sec: Exp. Methods}

We grow 6T films by thermal evaporation in an organic molecular beam deposition vacuum chamber equipped with a beryllium window for \textit{in situ} x-ray measurements at a base pressure of $7\cdot10^{-7}$\, mbar. Cleaved KCl substrates are heated up to $420\,^\circ$C in vacuum to reduce surface contamination prior to the deposition. The films are grown with molecular deposition rates between 1 and 1.5\,\AA/min at $50\,^\circ$C substrate temperature to a thickness of $150 \pm 20$\,\AA. The film thickness is monitored with a quartz crystal microbalance during growth. The grown thin films are analyzed by means of x-ray diffraction in a $\theta-2\theta$ geometry, in which the reflectivity in dependence of the out-of-plane scattering vector $q_{z} = \dfrac{4\pi}{\lambda}\sin \theta$ is studied at the corresponding values of the different crystal phases. The measurements are performed at a $Cu\,K_{\alpha}$ rotating anode system with a wavelength of $\lambda$=1.5406 \,\AA \, in nitrogen atmosphere. Using a temperature controlled stage we heat the substrate to evaluate molecular desorption from the decreasing intensity of the Bragg reflections of the LT and HT crystal phases. After reaching the desired temperature, we first monitor the x-ray reflection intensity of the HT phase at $q_z=0.907$\,\AA$^{-1}$ and subsequently the reflection of the LT phase at $q_z=0.838$\,\AA$^{-1}$.

\section{Relative Stability of HT and LT Polymorphs}
\label{Sec: Relative Stability of HT and LT}

The large LT unit cell is almost symmetrical with respect to a plane parallel to the $ab$ plane, as indicated in Fig. \ref{Fig: Crystal Structures}. Thus the unit cell can be approximately divided into two halves, with lattice constants $a$, $b$, and $c=22.35 \mbox{ \AA}$, containing two molecules each. The total energies of these smaller structures, obtained after minimizing the atomic forces, differ by less than 0.2 meV/molecule. Since this value is within our computational accuracy (see Sec. \ref{Sec: Theory}), we can consider the LT polymorph in such reduced unit cell \bibnote{We take the top half in Fig. \ref{Fig: Crystal Structures}. However, this choice does not affect what follows.}. This allows us to significantly decrease the computational costs, ensuring the same numercial accuracy for both polymorphs. In the remainder of this article the label LT will refer to the reduced unit cell. 

The relative stability of the HT and LT polymorphs is determined by their total energies. By employing the LDA functional, we find the LT phase to be more stable by 36 meV/molecule, compared to the HT phase. This energy difference increases to 51 meV/molecule, when we explicitly account for van der Waals interaction using the DFT-D2 method. This shows that both functionals lead to the same qualitative picture.

\begin{figure}
\centering
\includegraphics[width=.45\textwidth]{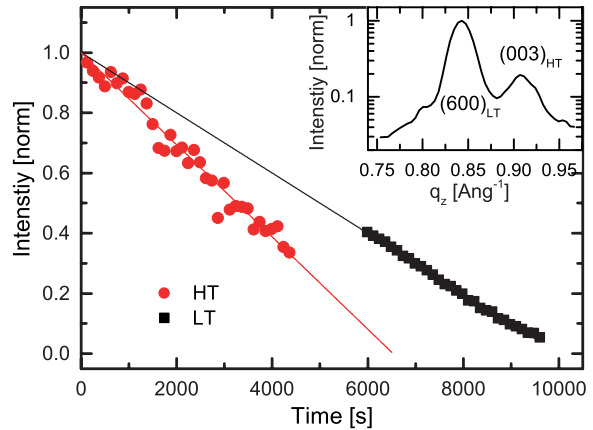}
\caption{(color online) Decay of the HT and LT Bragg reflection intensities over time at a substrate temperature \textit{T}=$428 \pm 5\, K$. Inset: $\Theta-2\Theta$ scan of the monitored LT and HT reflections. \label{Fig: ExperimentalData}}
\end{figure}

Experimentally, we determine the more stable phase, as well as the difference in the desorption energy barrier $E_d$ of the two crystal structures, by measuring the desorption rates of both LT and HT phase crystallites at fixed temperatures. A 15 nm thick, polycrystalline 6T thin film on NaCl exhibits phase coexistence as seen from the characteristic $(006)_{LT}$ and $(003)_{HT}$ Bragg reflections that both occur in a $\Theta -2\Theta$ scan (see Figure \ref{Fig: ExperimentalData}, inset). Heating this 6T film to a temperature of $428 \pm 5\, K$, corresponding to the onset of molecular desorption, we observe that the intensities of the Bragg reflections drop linearly as shown in Figure \ref{Fig: ExperimentalData} . Interestingly, the HT phase desorbs at a faster rate $[R^{HT} = (1.531 \pm 0.050) \cdot 10^{-}4$\, s$^{- 1}]$ than the LT phase 
$[R^{LT} = (1.000 \pm 0.011) \cdot 10^{-}4$\, s$^{- 1}]$, indicating a higher stability of the LT phase.

For a quantitative analysis, we explain the differences between the temporal decay of the two Bragg intensities by a difference in desorption energy barriers $E_d$.
Assuming that the Bragg intensity is directly proportional to the respective amount of HT or LT phase, the constant slope of the decay curves can be explained by molecular desorption from step edges at a constant rate without any significant morphological changes of HT and LT islands.
In atomic force microscopy measurements, resolving molecular terraces of standing upright molecules, we find no distinctly different HT and LT islands, so that a similar geometry is assumed for
both phases. We use an Arrhenius-type relation of the desorption rate $R = - A\, e^{- E_d/k_{B}T}$ with the molecular desorption energy $E_d$, the (constant) temperature \textit{T}, and an attempt frequency \textit{A}. Assuming
$A = A^{LT} = A^{HT}$ for both phases, one can write
\begin{eqnarray}
\ln\left(\frac{R^{HT}}{R^{LT}}\right)&=&-\frac{E_d^{HT}}{k_B T}+\frac{E_d^{LT}}{k_B T}.
\end{eqnarray}	
Therefore the desorption energy difference
$\Delta E_d = E_d^{LT} - E_d^{HT}$ is given by
\begin{eqnarray}
\Delta E=\ln\left(\frac{R^{HT}}{R^{LT}}\right)\cdot k_B T  .
\end{eqnarray}
From the decay rates we estimate $\Delta E_d = 15.7 \pm 3.1$ meV between the two phases. This finding is in qualitative agreement with theory, however, the energy difference is about 3 times smaller than the computed difference in relative stability. We can identify two possible sources of such discrepancy. Most important, the process of thermal desorption occurs at the surface and in particular at step edges and corners. In this case each molecule interacts only with a reduced number of nearest neighbors compared to the bulk, and therefore the resulting binding energy is intrinsically lower. Moreover, DFT calculations do not take into account thermodynamical effects. Although most of these contributions cancel out when considering energy differences, they still may lead to a systematic overestimation of the experimental values, as previously pointed out for other organic crystals \cite{nabo+08prb}. Overall, we claim good agreement between our theoretical and experimental results, which identify the LT phase as the more stable one. It is finally worth mentioning that our result is in contrast with a previous work based on classical force-fields calculations \cite{Della_2008_JPCA}. In that case, the authors found the HT polymorph to be energetically favored with respect to LT by about 15 meV/molecule. Although the absolute value of this difference is rather small, we can attribute the better accuracy and, importantly, the correct sign of our result to the inclusion of quantum effects.


\color{black}
\section{Transition path between HT and LT phases}
\label{Sec: Transition Path}
\subsection{Construction of the path}
\label{Sec: Construction of the Path}

\begin{figure*}
\centering
\includegraphics[width=0.9\textwidth]{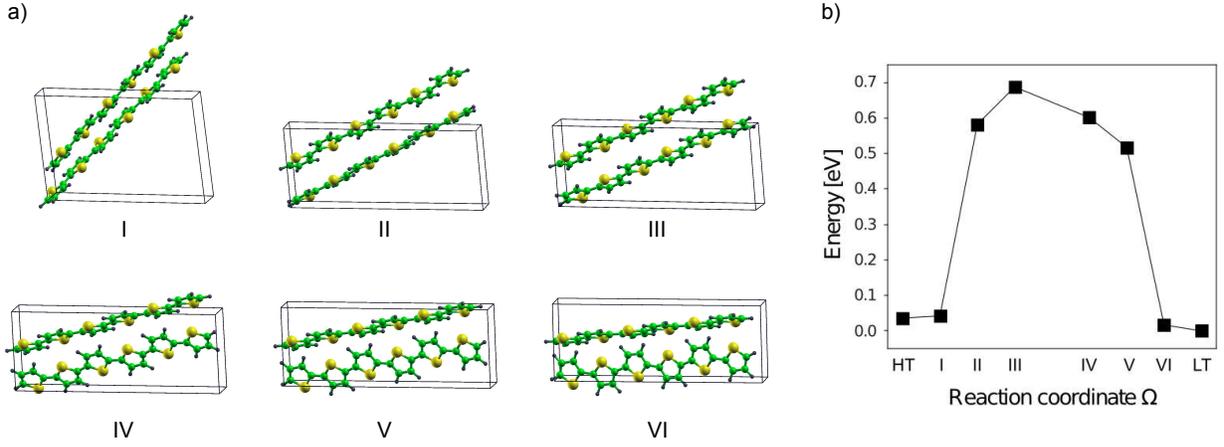}
\caption{(color online) a) Intermediate structures along the proposed transition path (see also Tab. \ref{Tab:Structural Parameters}). b) Energy barrier between HT and LT polymorphs of 6T. The reaction coordinate $\Omega$ is a function of the lattice parameters $a$, $b$, and $c$, as well as the angles $\phi$ and $\tau$.   \label{Fig: Intermediate-Structures_Barrier}}
\end{figure*}

In Sec. \ref{Sec: Relative Stability of HT and LT} we have clarified the higher relative stability of the LT polymorph with respect to the HT one. However, the energy difference between the two polymorphs is not the only factor that determines the material to grow in one phase rather than the other. One aspect is that film growth not only involves thermodynamical stability, but also kinetic processes. A quantity of interest in this context is the energy required to transform one structure into the other. To this extent we propose a possible transition path. Thereby, we face the challenge of determining the intermediate structures. While a variety of methods have been proposed and employed for such purpose \cite{McKe-Page1993JWL}, only a few of them can be applied to molecular crystals. In fact, in these systems molecules should keep their shape and be able to reorient themselves with respect to each other, while the unit cell adjusts accordingly. To fulfill these requirements, we adopt the so-called \textit{drag} method \cite{Henk-Joha-Jons2002Springer}. In our case, we treat molecules as rigid, while interpolating lattice parameters, as well as herringbone and tilt angles, the latter defining the orientation of the molecules with respect to each other and to the unit cell, respectively.

Six intermediate structures are constructed and shown in Fig. \ref{Fig: Intermediate-Structures_Barrier}a, labeled from \textrm{I} to \textrm{VI}. In Tab. \ref{Tab:Structural Parameters}, lattice constants, as well as herringbone and tilt angles are reported. For comparison, also the structural parameters of the HT and LT phase are shown, highlighted in bold. It is worth noting that interpolation of $a$, $b$, $c$, $\phi$ and $\tau$, indirectly determines the angles between the lattice parameters. Hence, while the initial HT and LT structures are monoclinic, the intermediate ones become triclinic. By inspecting Fig. \ref{Fig: Intermediate-Structures_Barrier}a the variation of herringbone and tilt angles is visible, as well as the change of the lattice parameters.
The six intermediate structures are optimized to minimize the atomic forces, while the unit cell parameters are held fixed to the values reported in Tab. \ref{Tab:Structural Parameters}. 

\subsection{Relative stabilities and electronic structure}
\label{Sec: Relative stabilities and electronic structure}

\begin{table}[h]
\setlength{\tabcolsep}{4pt}
\normalsize
\begin{tabular}{l | c | c | c | c | c | c | c | c }
 & \textbf{HT} & \textrm{I} & \textrm{II} & \textrm{III} & \textrm{IV} & \textrm{V} & \textrm{VI} & \textbf{LT}\\
\hline \hline
LDA & \textbf{36.2} & 42.2 & 582.3 & 687.4 & 601.1 & 517.4 & 16.8 & \textbf{0.0} \\
\hline 
vdW & \textbf{51.0} &  &  & &  &  &  & \textbf{0.0} \\
\hline
\end{tabular}
\caption{Energy difference per molecule of each structure with respect to the LT phase, as obtained from LDA and DFT-D2 (labeled vdW). 
All energies are given in meV/molecule.\label{Tab: Relative Energies}}
\end{table}

As a next step, we evaluate the total energies of the optimized intermediate structures. 
In this way, we are able to estimate the energy barrier between the HT and LT phases.
The results of these calculations are reported in Tab. \ref{Tab: Relative Energies} and in Fig. \ref{Fig: Intermediate-Structures_Barrier}b. The relative energies with respect to the most stable polymorph, the LT phase, are displayed as a function of the reaction coordinate $\Omega$, which is defined by the lattice parameters and the angles $\tau$ and $\phi$. 
Overall the barrier presents a \textit{top hat} shape. While structure \textrm{I} resembles the HT phase and is energetically very close to it, structure \textrm{VI} is similar in energy and shape to the LT phase (see Tab. \ref{Tab: Relative Energies}). On the other hand, structures \textrm{II}-\textrm{V} are significantly less favored, exhibiting energies more than 0.5 eV/molecule higher compared to LT. The most unfavored polymorph is structure \textrm{III}. The total energy exceeds that of the LT phase by about 0.7 eV per molecule, which represents the apex of the dome. Considering an increase in energy by explicitely accounting for van der Waals interactions, we estimate the upper bound of the barrier to be about 1 eV.
Similar results have been obtained for other organic crystals, such as \textit{para}-sexiphenyl on a mica step-edge \cite{Hlaw+2008SCI}. \newline

\begin{figure*}
\centering
\includegraphics[width=0.99\textwidth]{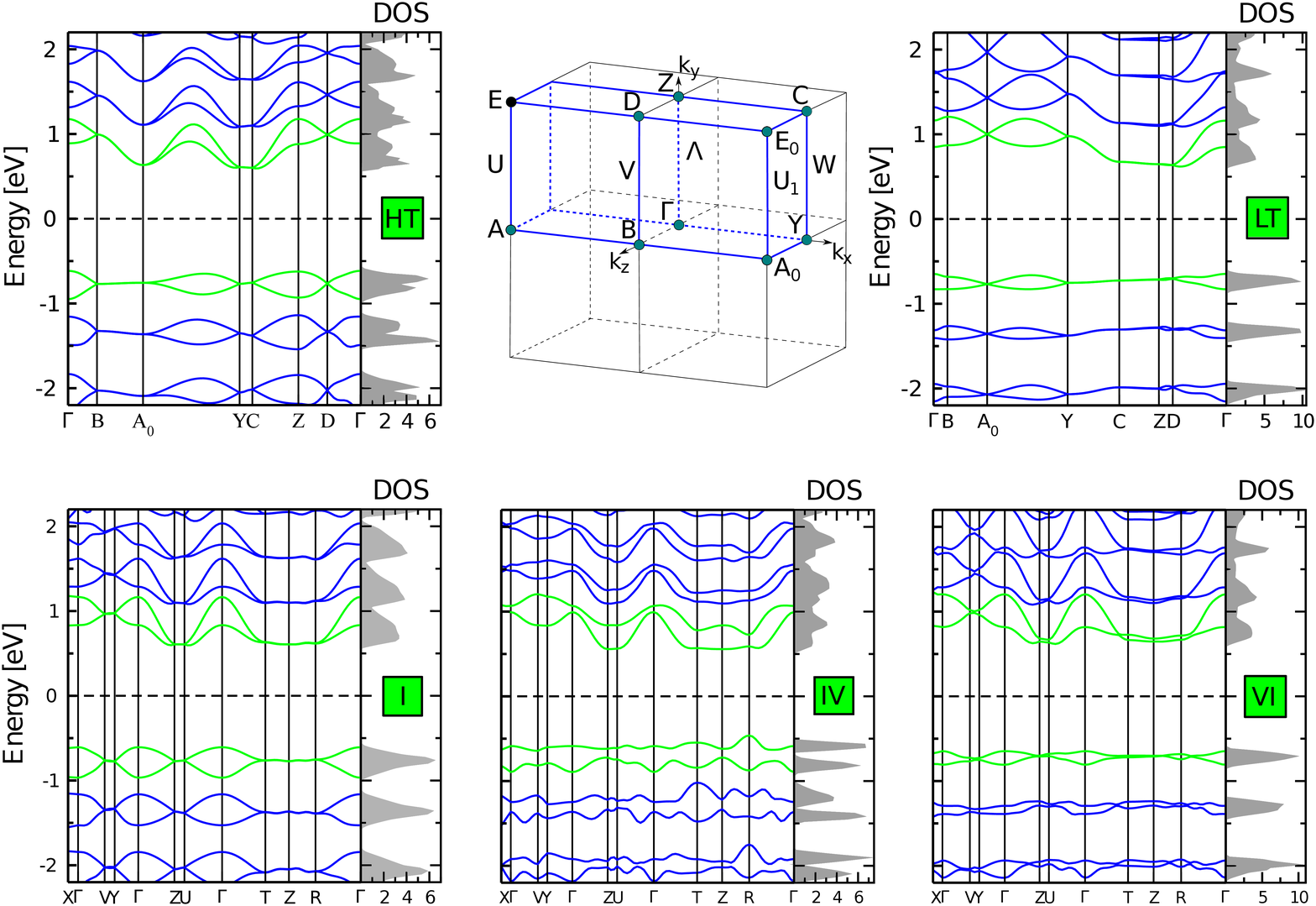}
\caption{(color online) Band structure and density of states (DOS) of HT and LT, as well as selected intermediate structures (see Fig. \ref{Fig: Intermediate-Structures_Barrier}).  The subbands corresponding to the uppermost valence-band (VB) pair and lowest conduction-band (CB) pair are highlighted in green. The Brillouin zone of HT and LT is shown in the middle of the top row. The DOS are expressed in 100 states/eV. The Fermi energy is set to 0 eV.
\label{Fig: All-Bands}}
\end{figure*}

We finally present, in Fig. \ref{Fig: All-Bands}, the electronic properties of selected intermediate structures, compared to the HT and LT ones. These results allow us to further characterize the predicted metastable structures. 
In Fig. \ref{Fig: All-Bands} the band structure and density of states (DOS) of the three selected intermediates \textrm{I}, \textrm{IV} and \textrm{VI} are shown in comparison with the stable polymorphs HT and LT. We notice the typical features of organic crystals (see e.g. Ref. [\onlinecite{Hummer_Draxl_ElectronicProperties}]). In both the valence and conduction regions, subbands are arranged in pairs, according to the double multiplicity of the 6T molecules in the unit cell. These features are reflected also in the DOS, which presents in both cases relatively sharp peaks in the valence region, corresponding to the different subbands. 
The subbands in the conduction region are energetically closer to each other, and present overall increased dispersion compared to the valence region. The highest valence-bands (VB) and lowest conduction-bands (CB) are highlighted in green. Both polymorphs have indirect Kohn-Sham (KS) band gaps of 1.2 and 1.1 eV for LT and HT, respectively. 

In the LT phase the VB bandwidth is 0.2 eV, while it is larger (0.5 eV) in the CB. For symmetry reasons, both bands are degenerate along the path from $Y$ to $C$, and exhibit a small splitting between $C$ and $Z$. The splitting is largest at the $\Gamma$-point and halfway inbetween the points $A_0$ and $Y$. In the HT phase, the bandwidth is twice as large in the VB (0.4 eV), and slightly larger in the CB (0.6 eV), compared to LT. The largest splitting is found at $\Gamma$ and $Z$, as well as halfway between $A_0$ and $Y$. Bands are degenerate between $Y$ and $C$, as well as between $B$ and $A_0$. 

For comparison, we have selected those structures (\textrm{I}, \textrm{IV} and \textrm{VI}), which mostly differ from each other in the arrangement of the molecules and in the energetics (see Figs. \ref{Fig: Intermediate-Structures_Barrier}a and b). These intermediate structures belong to the space group $P1$, exhibiting trivial symmetry. These systems present indirect KS band gaps of 1.0 eV (\textrm{I}), 0.9 eV (\textrm{IV}), and 1.2 eV (\textrm{VI}). These values are comparable to those of the HT
(1.1 eV) and LT (1.2 eV) phases. The lower stabilities of these systems with respect to the stable HT and LT polymorphs is evident from the band structures. In particular, in structure \textrm{IV}, which is clearly unfavored in the chosen transition path, a large band splitting is observed, in both valence and conduction regions. Especially the occupied states feature sharp peaks in the DOS. This is a signature of the lower symmetry of this system, compared to the stable HT and LT polymorphs. The other intermediate structures, \textrm{I} and \textrm{VI}, which are structurally and energetically close to HT and LT, respectively, still present subbands, and the fingerprints of reduced symmetry, such as subband degeneracy, are less pronounced. Overall, the bands exhibit low dispersion, especially for the \textrm{IV} and \textrm{VI} structures. This implies high effective carrier masses and therefore low charge-carrier mobilities in these regions. The bandwidths in the VBs decrease from 0.4 eV for \textrm{I} and \textrm{IV} to 0.2 eV in structure \textrm{VI}. Thus, structure \textrm{I} (\textrm{VI}) shares the same bandwidth in the VB as the HT (LT) polymorph. The CB bandwidths tend to be larger with values of 0.6 eV (structure \textrm{I}), 0.7 eV (structure \textrm{IV}) and 0.7 eV (structure \textrm{VI}), thus they are 0.1 eV larger than those of the HT and LT phase, respectively. 

\section{Summary and Conclusions}
\label{Sec: Summary}
We have presented a combined theoretical and experimental study on polymorphism in 6T crystals. We have clarified that the LT phase is favored over the HT one by about 50 meV/molecule, as obtained from DFT calculations, explicitly taking into account van der Waals interactions. This result is in agreement with our thermal desorption measurements. Our results confirm the importance of explicitly accounting for quantum effects and dispersive intermolecular interactions, to quantify the relative stability of different polymorphs in organic crystal structures.

In addition, we have proposed a transition path between the two stable 6T polymorphs, estimating the energy barrier between the HT and LT phase of about 1 eV/molecule. The results are supplemented by a thorough analysis of the electronic properties of the stable and selected intermediate structures.

\section*{Acknowledgement}
We acknowledge fruitful discussions with Hong Li, Dmitrii Nabok, and Peter Sch\"afer. 
This work was partly funded by the German Research Foundation (DFG) through the Collaborative 
Research Centers SFB 658 and SFB 951. C.~C.~ acknowledges support from the Berliner Chancengleichheitsprogramm (BCP). L.~P.~ acknowledges support from the Studienstiftung des deutschen Volkes.

\end{document}